\documentclass[aps,twocolumn,prl]{revtex4}
\usepackage{amsfonts}
\usepackage[final,hiresbb]{graphicx}
\usepackage{amsmath}
\usepackage{amssymb}
\usepackage{amstext}
\usepackage{subfigure}
\usepackage{color}

\begin{document}

\title{A Two-Dimensional Carbon Semiconductor}
\author{David J. Appelhans, Zhibin Lin and Mark T. Lusk}
\affiliation{Department of Physics, Colorado School of Mines, Golden, CO 80401, USA}

\begin{abstract}
We show that patterned defects can be used to disrupt the sub-lattice symmetry of graphene so as to open up a band gap.  This way of modifying graphene's electronic structure does not rely on external agencies, the addition of new elements or special boundaries. The method is used to predict a planar, low energy, graphene allotrope with a band gap of 1.2 eV. This defect engineering also allows semiconducting ribbons of carbon to be fabricated within graphene. Linear arrangements of defects lead to naturally embedded ribbons of the semiconducting material in graphene, offering the prospect of two-dimensional circuit logic composed entirely of carbon.
\end{abstract}

\maketitle

\begin{figure}[hptb]\begin{center}
\includegraphics[width=0.48\textwidth]{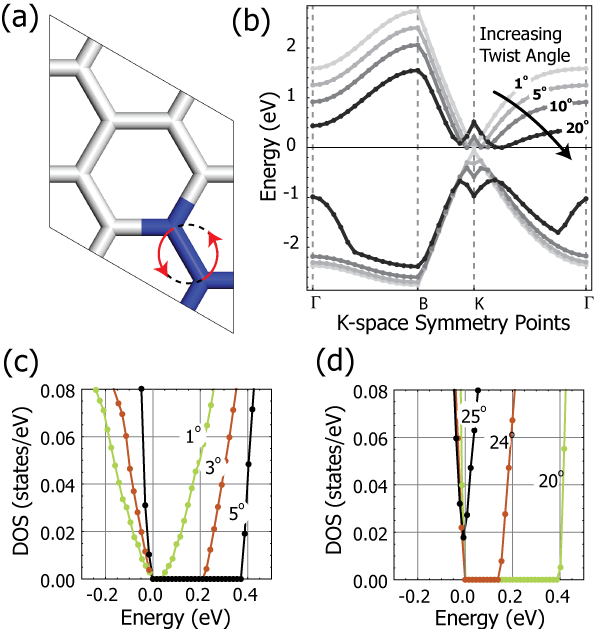}
\caption{%
The twisting of a bond within a periodic cell (a) disrupts lattice symmetry, causing band gap opening (b,c) and closing (d). The Fermi energies have been aligned to more clearly see the effect of bond rotation.}\label{BondRotation}%
\end{center}
\end{figure}

Graphene-based electronics promise to overcome limitations associated with silicon technologies~\cite{ref:Novoselov2007,ref:geim-the-rise-of-graphene,ref:Bolotin2008,ref:Lin2010} and open previously unavailable new applications~\cite{ref:Schedin2007}. The improved performance derives from unparalleled strength~\cite{ref:Lee2008} combined with high electron mobility~\cite{ref:Novoselov2005nature,ref:Bolotin2008}. Graphene is a semi-metal, though, and therefore requires modification to open up the band gap desirable for logic circuits. This is simple to do in principle, as demonstrated in Figure~\ref{BondRotation}, where small rotations of a bond, periodically repeated through the lattice, collapses the Dirac cones at the graphene K point.  Continued rotation of the bond will close the band gap elsewhere resulting in a conductor. To exploit this sensitivity, an agency must be found for tuning the distortion of lattice symmetry.

A number of approaches have been considered for distorting graphene so that it exhibits a band gap. All of these rely on either the action of an external agency or a confining boundary to disrupt the delicate sub-lattice symmetry. These include the use of strongly interacting substrates~\cite{ref:Ohta2006,ref:Giovannetti2007,ref:Ciobanu2009} or the application of electric fields in graphene bi-layers~\cite{ref:Ohta2006,ref:zhang2005,ref:Nilsson2007}. Quantum confinement effects can also be exploited as in carbon nano-ribbons~\cite{ref:Brey2006, ref:Mucciolo2009, ref:Hod2007, ref:Barone2006}, or carbon nano-meshes~\cite{ref:Bai2010}. The graphene can also be chemically modified. For instance, hydrogen can be introduced so as to change the bonding from sp-2 to sp-3 to create the corrugated insulator graphane~\cite{ref:Sofo2007}. Similarly, a graphene oxide insulator can be created via bonding with hydroxyl groups~\cite{ref:Dikin2007}. The external agency could also be in the form of neighboring regions of foreign material as in composite two-dimensional structures that harbor domains of graphene~\cite{ref:Ci2010}.

\begin{figure}[hptb]\begin{center}
\includegraphics[width=0.48\textwidth]{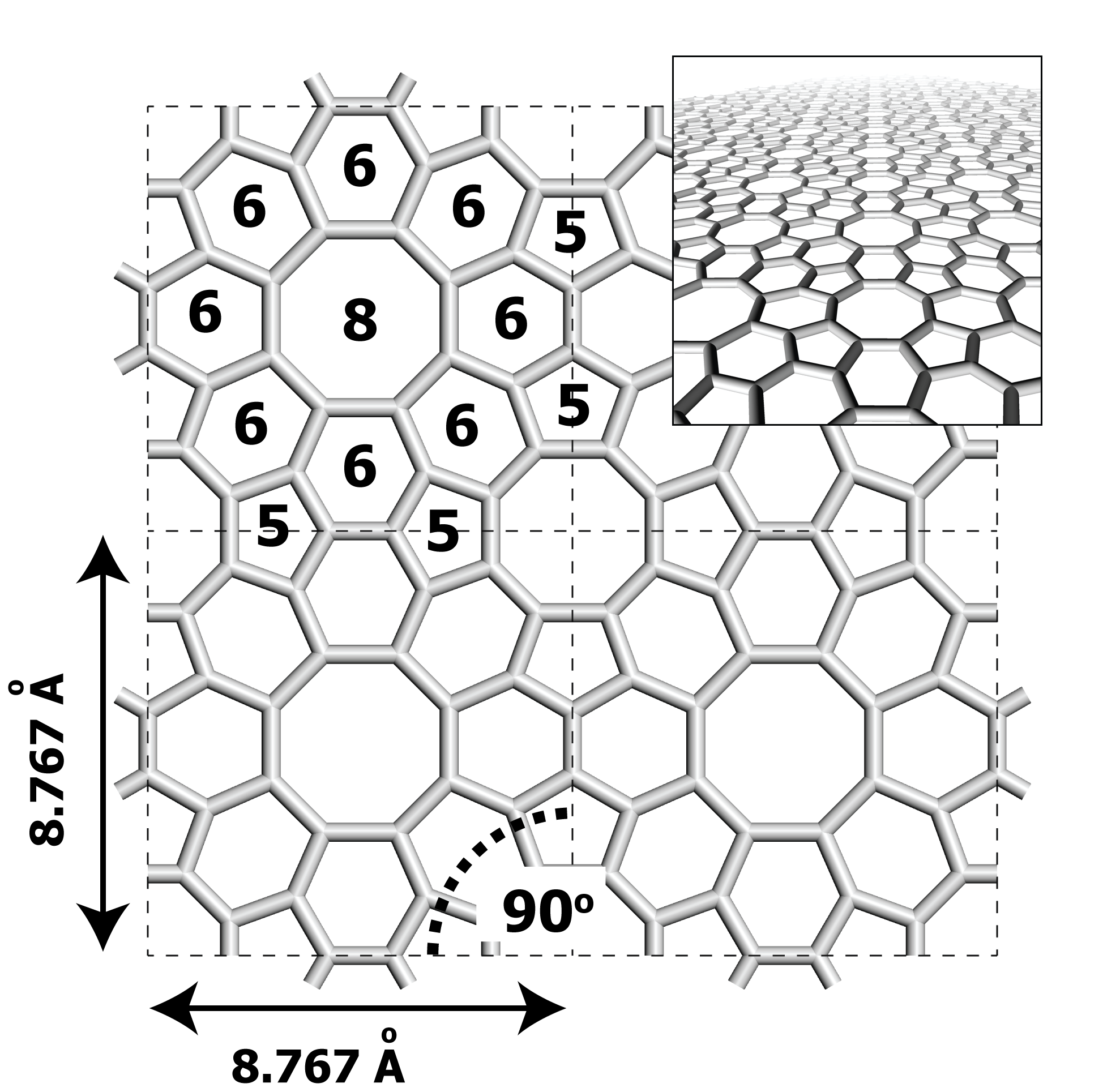}
\caption{%
A computationally created semiconducting allotrope of graphene resulting from patterned defects. The planar structure is only 313 meV/atom above graphene, comparable to 233 meV/atom for Haeckelite H567, a metallic graphene allotrope previously theorized~\cite{ref:Terrones2000}.}\label{SCgeometry}%
\end{center}
\end{figure}

Here we explore an alternative approach wherein the symmetry of the graphene is modified using patterned defects. Previous defect investigations have predicted only metallic allotropes of graphene~\cite{ref:Crespi1996, ref:Terrones2000, ref:LuskCarr2009, ref:AppelhansThesis}. Interestingly, all of these conducting allotropes can be constructed by patterning two types of defects~\cite{ref:LuskCarr2009}, Stone-Thrower-Wales (STW)~\cite{ref:stone1986,ref:thrower1978} and Inverse Stone-Thrower-Wales (ISTW)~\cite{ref:LuskCarr2008, ref:LuskCarr2010} defects. The consistent metallic nature of these graphene allotropes, though, suggests that their local lattice distortions are not appropriate for delicately teasing apart the K point Dirac cones without pinching off the band gap elsewhere.  This motivated the search for a new defect which might produce semiconducting graphene, and we have discovered that the di-vacancy (DV) defect is well-suited for the task. A patterned combination of DV and STW defects results in a planar structure with a band gap, a semi-conducting form of graphene. This stable, single-layer, all carbon semiconductor, named \emph{Octite SC}, is shown in Figure~\ref{SCgeometry}. We recently became aware of a complimentary approach in which pentagon and heptagon defect were coupled with surface corrugation to introduce a band gap in graphene~\cite{ref:Nunes2010}.

As a starting point, Density Functional Theory (DFT) was employed to estimate the ground state structure and predict electronic character. A real-space numerical atomic orbital code~\cite{ref:Delley1990} was used to initially relax the structures until the energy change was less than $2.7\times10^{-4}$ eV. The Perdew-Wang generalized gradient approximation accounted for electron exchange and correlation energy~\cite{ref:Perdew1992}. This computational setting has been shown to accurately predict defect geometries and energies associated with graphene structures~\cite{ref:LuskCarr2008,ref:LuskCarr2009,ref:LuskCarr2010}.

The new allotrope has a primitive cell with 28 atoms composed in pentagons, hexagons, and octagons. It has a planar density of 0.364~atoms per $\AA^2$, comparable to 0.380~atoms per $\AA^2$ for graphene. The central geometrical feature is an octagon completely surrounded by hexagons. The cell geometry is square with p4/MMM symmetry. A Hessian linear vibrational analysis showed the structure to be a local energy minimum. Room temperature quantum molecular dynamics simulations were also used to confirm its stability.

As a final check on the ground state structure, the cell geometry and linear stability were verified using a second DFT implementation that employs a plane-wave basis set~\cite{VASP3, VASP4}. A Projector Augmented Wave approach~\cite{VASPpaw1} was used within a Generalized Gradient Approximation~\cite{VASP-GGA1}. The wave function energy cutoff was set at 400 eV, and a conjugate gradient method employing a 4x4x2 Monkhorst-Pack grid resulted in the same ground state structure but with a lattice constant of 8.674 \AA.

\begin{figure}[ptb]\begin{center}
\includegraphics[width=0.45\textwidth]{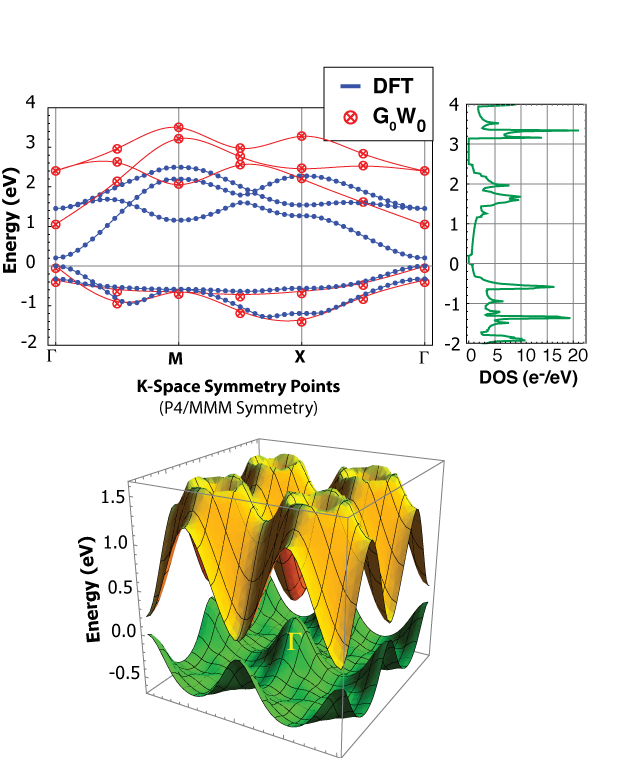}
\caption{Band structure (BS) and density of states (DOS) for a semiconducting graphene allotrope. Both DFT and $G_0W_0$ predictions are shown. The DOS plot at right is from the DFT calculation.}
\label{BDandDOS}
\end{center}\end{figure}

The plane-wave DFT code was subsequently used to predict both the band structure (BS) and Density of States (DOS). A 24x24x1, $\Gamma$-centered k-space grid was used to calculate the electronic properties. The BS and DOS diagrams are shown in Figure~\ref{BDandDOS}. The material possesses a 0.2 eV direct band gap at the $\Gamma$ point. As in graphene, all carbon atoms bond with three nearest neighbors, but the symmetry of the in-plane sp-2 character has been disrupted by the varying bond angles and bond lengths.

DFT typically underestimates the band gap of periodic structures~\cite{ref:payne-iterative-minimization}, so the 0.2 eV gap should be viewed as a lower bound. Many-body perturbation theory, within the $G_0W_0$ approximation~\cite{Shishkin2007}, was therefore used to provide a more accurate estimate of the electronic structure. As expected, the inclusion of quasi-particle effects widens the band gap, providing an estimate for the direct, $\Gamma$ point gap of 1.1 eV. The $G_0W_0$ band structure is shown with its DFT counterpart in Figure~\ref{BDandDOS}. Three iterations towards self-consistency ($GW_0$ approximation) opened the band gap slightly to 1.2 eV.

Having elucidated the lattice and electronic structure of this new material, we now examine its defect architecture. Such decompositions are not intended as  prescriptive synthesis steps and are, rather, a way of understanding the properties of graphene allotropes from the broader perspective of defects from which they are composed. All Haeckelite allotropes can be constructed from graphene with templates of either STW defects or a combination of STW and ISTW defects~\cite{ref:LuskCarr2009}. Likewise dimerites, as the name suggests, amount to graphene on which a regular grid of ISTW defects have been applied~\cite{ref:LuskCarr2009}. The new semiconducting allotrope, however, requires the addition of di-vacancies (DV) to the defect alphabet. By patterned placement of DVs and STW defects, Octite SC can be engineered from graphene.

\begin{figure}[ptb]\begin{center}
\includegraphics[width=0.45\textwidth]{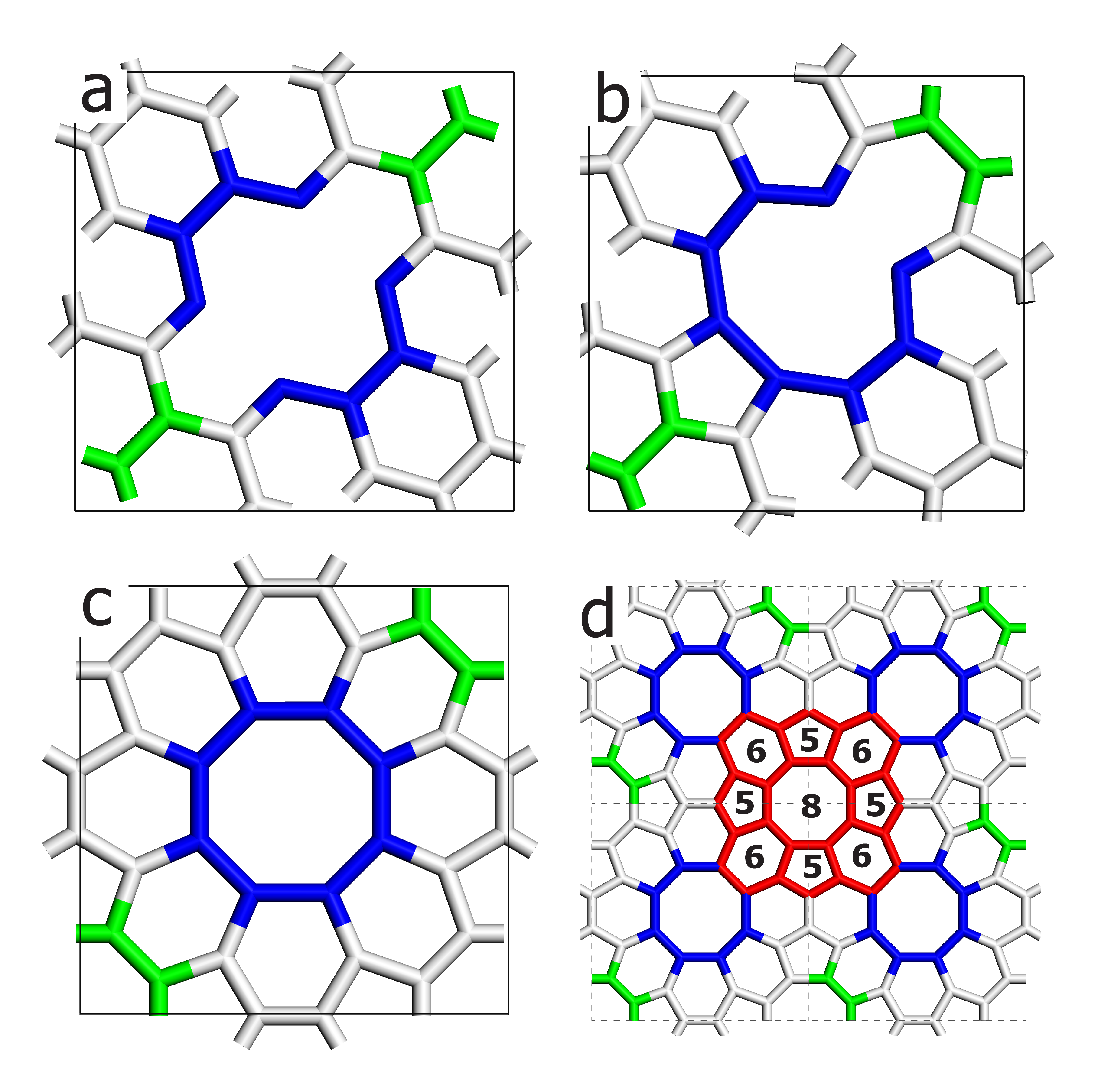}
\caption{\label{Synthesis}Patterned defect composition of Octite SC from graphene. (a) Graphene with a di-vacancy. (b) A STW defect is introduced by rotating the gray (green) bond. (c) A second STW defect is introduced. (d) Patterning of the patch results in a second octagonal defect surrounded by alternating hexagons and pentagons, shown in gray (red).}
\end{center}\end{figure}

We schematically illustrate the defect composition in Figure~\ref{Synthesis}. Initially a DV defect is introduced as shown in Figure~\ref{Synthesis}(a).  Two bond rotation defects are subsequently imposed (frames b and c). The result is then replicated so as to create an array of octagonal defects. Octagons surrounded by hexagons result from DV and STW defects while smaller octagons surrounded by alternating pentagons and hexagons derive from STW defects (frame d).

The formation energy associated with each defect shown in Figure~\ref{Synthesis} was calculated within a periodic 200 atom supercell as well as on a passivated, 206-atom flexible graphene flake. The formation energy of the DV was found to be 7.6 eV in the periodic supercell, in good agreement with previous periodic DFT estimates of 7.5 eV~\cite{ref:Kotakoski2006} and 8.7 eV~\cite{ref:ElBarbary2003}. The flexible boundaries of the graphene flake allowed the formation energy to be substantially lower at 6.6 eV. We calculate that the subsequent bond rotation defects on the right and left of the DV result in sequential, net increases of 4.7 eV and 5.1 eV for the periodic supercell or 5.0 eV and 5.1 eV for the graphene flake.

DV defects, central to the structure of Octite SC, can be precisely created via irradiation~\cite{ref:Hashimoto2004}. Recently, an electron irradiation beam focused to $1~\AA$ was shown to be capable of precisely creating vacancies in CNTs~\cite{ref:Banhart2009}. The stability of DVs has also been considered theoretically within a CNT setting, where molecular dynamics annealing predicted that they will remain stable in the 5-8-5 configuration up to at least 2700 K~\cite{ref:Yuan2009} and 3000 K~\cite{ref:Sammalkorpi2004}. Tight-binding molecular dynamics studies at higher temperatures concluded that DVs were stable out to 3100K for at least the 90 ps computational experiment~\cite{ref:Lee2005}. A subsequent temperature increase to 3800 K was required to cause this 5-8-5 defect to relax to the slightly lower energy 555-777 defect with an estimated barrier of 5.2 eV~\cite{ref:Lee2005}. Our own transition state analysis estimated a similar barrier of 3.7 eV for relaxation from a 6-8-6 defect to a 555-777 defect. These results suggest that the DV defect, central to Octite SC, should be stable for the operating conditions expected of standard electronic circuitry.

STW defects have been extensively studied in both graphene~\cite{ref:Li2005,ref:stone1986} and CNTs~\cite{ref:Kotakoski2006,ref:Kim2006}. STW defects could be created or reversed by scanning tunneling microscopy (STM)~\cite{ref:berthe2007} or atomic force microscopy (AFM)~\cite{ref:sugimoto2005}. Both DV and STW defects have been experimentally observed~\cite{ref:Banhart2009,ref:Hashimoto2004,ref:Urita2005}.

\begin{figure}[ptb]\begin{center}
\includegraphics[width=0.3\textwidth]{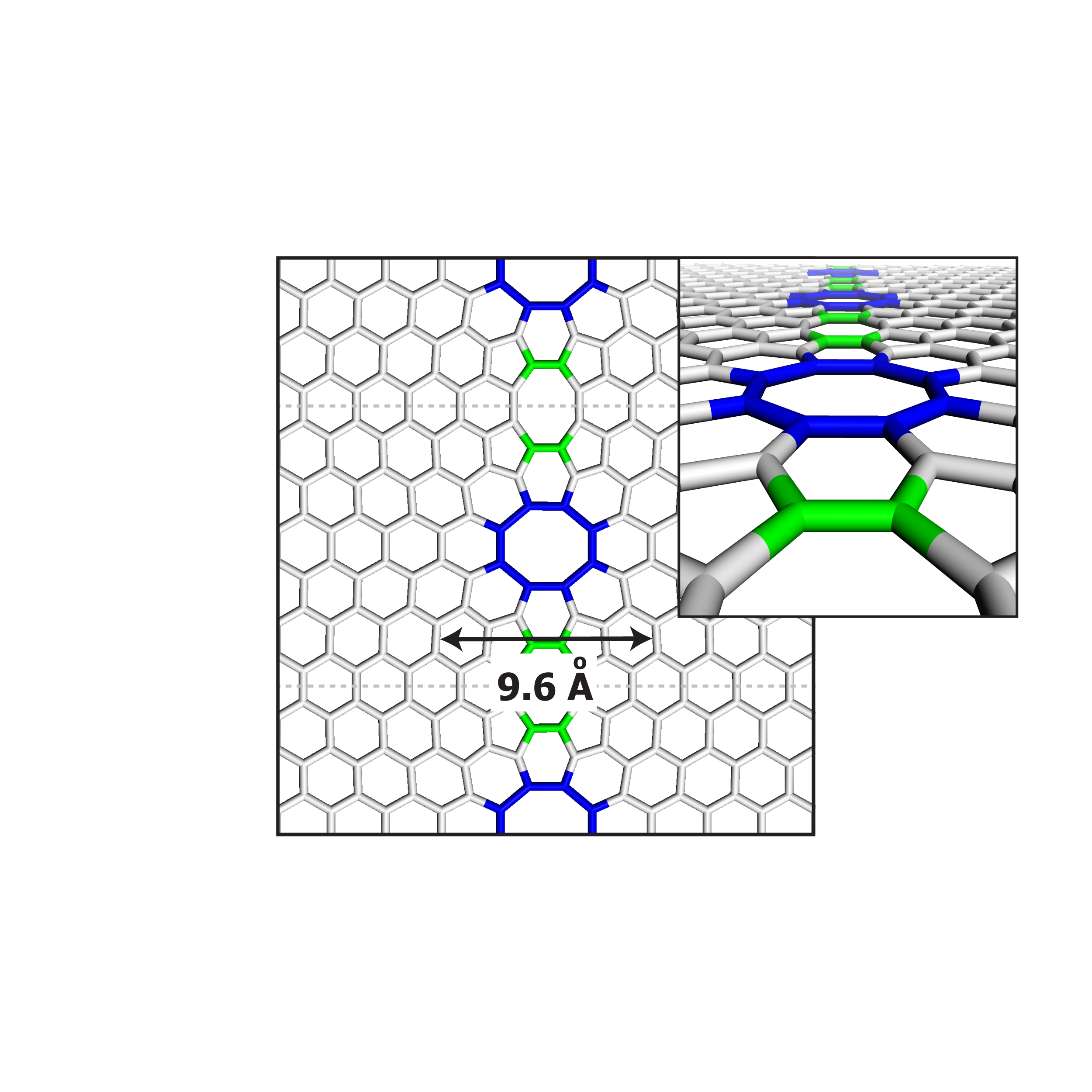}
\caption{\label{Ribbon}A ribbon of Octite SC built within graphene using DV and STW defects.}
\end{center}\end{figure}

Semiconducting ribbons can be built into graphene as shown in Figure~\ref{Ribbon}. Our DFT calculations predict that these ribbons remain planar within the graphene, and preliminary results suggest that the semiconducting character is locally preserved. Analogous conducting structures can also be designed, and our prediction of a metallic defect ribbon has already been experimentally verified~\cite{ref:Batzill2010,ref:AppelhansThesis}. Semiconducting and conducting ribbons, embedded within graphene, could be used together as the basis for electronic logic circuits.

In summary, we have found that patterned defects involving DVs yield a planar, single-atomic layer material composed entirely of carbon that is an intrinsic semiconductor; no external agencies or special boundary conditions need to be imposed in order to open up a band gap. In comparison with epitaxial paradigms, this is particularly attractive since suspended graphene possesses a mobility ten times that of graphene fabricated on a substrate~\cite{ref:Bolotin2008}. Future electronics designs can exploit local and extended defects to build free standing, single-atomic layer logic circuits based on the local symmetry properties of carbon bonds.

\section{Acknowledgements}
We are pleased to acknowledge the use of computing resources provided through the Golden Energy Computing Organization (NSF Grant No. CNS-0722415) and the Renewable Energy MRSEC program (NSF Grant No. DMR-0820518) at the Colorado School of Mines.


\end{document}